# Report from Workshop on Dialogue alongside Artificial Intelligence

Thomas J McKenna (Boston University), Ingvill Rasmussen (University of Oslo), Sten Ludvigsen (University of Oslo), Avivit Arvatz (The Hebrew University of Jerusalem), Christa Asterhan (The Hebrew University of Jerusalem), Gaowei Chen (The University of Hong Kong), Julie Cohen (University of Virginia), Michele Flammia (Independent Scholar), Dongkeun Han (University of Cambridge), Emma Hayward (University of Cambridge), Heather Hill (Harvard University), Yifat Kolikant (The Hebrew University of Jerusalem), Helen Lehndorf (Freie Universität Berlin), Kexin Li (The University of Hong Kong), Lindsay Clare Matsumura (University of Pittsburgh), Henrik Tjønn (University of Oslo), Pengjin Wang (The University of Hong Kong), Rupert Wegerif (University of Cambridge)

**November, 2025**

## Executive Summary

Educational dialogue – the collaborative exchange of ideas through talk – is widely recognized as a catalyst for deeper learning and critical thinking in and across context. At the same time, artificial intelligence (AI) has rapidly emerged as a powerful force in education, with the potential to address major challenges, personalize learning, and innovate teaching practices. However, these advances come with significant risks: rapid AI development can undermine human agency, exacerbate inequities, and outpace our capacity to guide its use with sound policy. Human learning presupposes cognitive efforts and social interaction (dialogues). In response to this evolving landscape, an international workshop titled "Educational Dialogue: Moving Thinking Forward"[1] convened 19 leading researchers from 11 countries in Cambridge (September 1–3, 2025) to examine the intersection of AI and educational dialogue. This AI-focused strand of the workshop centered on three critical questions: (1) When is AI truly useful in education, and when might it merely replace human effort at the expense of learning? (2) Under what conditions can AI use lead to better dialogic teaching and learning? (3) Does the AI–human partnership risk outpacing and displacing human educational work, and what are the implications? These questions framed two days of presentations and structured dialogue among participants.

---

[1] https://www.educ.cam.ac.uk/research/groups/cedir/events/Cambridge%20workshop%20Sept%202025/Call%20for%20applications%20to%20participate.pdf

**Report from Workshop on Dialogue alongside Artificial Intelligence**

# 1. Introduction

Key findings from the workshop highlight a fundamental tension: AI offers transformative opportunities to enhance educational dialogue, yet it also poses risks of undermining the authentic human interaction at the heart of learning. Participants showcased cutting-edge tools and empirical studies demonstrating AI's ability to visualize classroom discussions, provide real-time feedback, and scaffold productive talk. For example, recent classroom trials have shown that AI-based analytics and visualization tools can significantly improve teachers' dialogic practices and even boost student learning outcomes. Students themselves report that AI chatbots can serve as valuable 24/7 tutors – clarifying complex concepts, generating practice problems, and giving feedback on writing – thus supporting their study process. At the same time, both the workshop participants and emerging research caution against over-reliance on AI. If students or teachers become too dependent on AI-generated answers, it may diminish critical thinking, creativity, and metacognitive skills. Student surveys likewise reveal fears of losing independent problem-solving skills and "the human element" in learning due to AI. In short, the workshop underscored that AI works best as a mediating tool within human-centered tools, networks and infrastructures – not as a replacement for teachers or dialogue partners. This blended approach aligns with the broader consensus in AI-in-education research that AI should augment rather than supplant human instruction. Ensuring a human-centered, equitable integration of AI is crucial to harness its benefits while preserving the collaborative, inclusive nature of education.

In the sections that follow, we summarize the workshop's insights in a format akin to formal academic meeting proceedings. We first describe innovative frameworks, AI tools and empirical findings presented. We highlight critical issues and pedagogical tensions that emerged, from questions of agency and equity to concerns about authenticity of dialogue. We then review new methodological approaches and design principles for AI in educational dialogue. Finally, we distill future research directions and collaborative opportunities identified by the group, and we conclude with the broader implications of these discussions for the field of AI in Education. But first we start with an account of the methods used in generating the text.

## 2. Methods used in generating the text.

The meeting notes presented in this document were generated through a systematic computational approach that synthesized exclusively human-authored content. All substantive material included in this document derives directly from text written and shared by workshop participants through their application materials, presentation notes, slide decks, discussion contributions, and other workshop documentation. No artificial intelligence-generated content forms part of the corpus; rather, large language model (LLM) technology, specifically Claude Sonnet 4, was employed solely as a synthesis and organization tool to structure and present the human-authored materials in a coherent format.

**Data Sources and Input Materials**

The synthesis process drew exclusively from human-created primary source documents from workshop participants, including: (1) individual participant application materials containing research backgrounds and proposed contributions; (2) technical documentation and white papers authored by participants describing AI systems and methodologies; (3) presentation materials, slide decks, and notes created by participants; (4) transcribed discussion contributions and workshop dialogue; and (5) structural guidelines provided by the workshop organizers regarding desired academic formatting and content organization. All theoretical insights, empirical findings, technical details, and methodological descriptions presented in these notes originate from the participants' own written and spoken contributions.

**Computational Processing Framework**

The initial text generation employed a multi-stage prompting methodology designed to extract, synthesize, and organize information from the source materials. The LLM was instructed to: identify key theoretical frameworks and empirical findings across participants' work; synthesize common themes and methodological approaches while preserving individual contributions; maintain academic objectivity and appropriate attribution; and structure content according to established academic meeting note conventions. The model utilized its training on academic discourse patterns to generate appropriate scholarly language, citation formats, and organizational structures.

**Human-AI Collaborative Refinement Process**

Following initial generation, the document underwent multiple iterative refinement cycles incorporating human feedback on: participant representation balance to ensure equitable coverage across all attendees; factual accuracy verification against source materials; academic tone and flow optimization for scholarly publication; structural coherence and logical progression of ideas; and removal of computational artifacts such as excessive bullet points and formulaic language patterns.

**Quality Assurance and Validation**

The final document represents the convergence of automated synthesis capabilities with domain expertise validation. Human oversight ensured that: theoretical concepts were accurately represented and appropriately contextualized; technical details of AI systems were precisely documented; participant contributions were proportionally and fairly attributed; and the overall narrative maintained academic integrity while capturing the collaborative nature of the workshop discussions.

## 3. Innovative Frameworks for Understanding and Analyzing AI in an Educational Context

An intriguing theoretical tool highlighted at the workshop was the DIKW pyramid – which stands for Data, Information, Knowledge, Wisdom – and its application to understanding AI's role in education. Pengjin Wang and Gaowei Chen (University of Hong Kong) demonstrated how this classic framework from information science can be adapted to educational settings shaped by AI. In their presentation, they discussed how raw classroom data (such as video recordings of lessons) can be transformed by AI into meaningful information (through processes like transcript coding and visualization), which teachers can then interpret to build knowledge about their practice, and ultimately, develop wisdom in pedagogical decision-making. By mapping this DIKW hierarchy onto real classroom use cases, Wang and Chen illustrated the journey from AI-processed data to teacher insights.

One example was their Classroom Discourse Analyzer (CDA), an AI-driven tool that processes classroom videos and generates visual representations of discussion patterns. The CDA automatically codes features of teacher–student dialogue (e.g. types of questions, talk turns, wait times) and presents these analyses in an accessible visual dashboard. This constitutes the "data-to-information" step: AI helps reduce the complexity of raw classroom data into interpretable charts and graphs. Teachers using these AI-generated visualizations can more easily recognize, for instance, whether they are guiding discussions toward Academically Productive Talk (APT) moves – such as pressing students to explain reasoning or building on each other's ideas – which are known to foster deeper understanding. Over time, engaging with such feedback allows teachers to calibrate their practice (closing the gap between current discourse patterns and ideal dialogic strategies), thereby turning information into actionable knowledge. With continuous reflection and collaborative discussion (e.g. in professional learning communities), teachers may even reach the "wisdom" level – developing an intuitive, value-oriented sense of how and when to use dialogic techniques for maximum student benefit.

Empirical evidence shared by Chen and colleagues strongly supports the value of this AI-augmented reflection process. In fact, they reported results from randomized controlled trials indicating that video-based professional development enhanced with AI analytics can significantly improve teaching quality and student learning outcomes. In one study, teachers who used the CDA visual feedback showed greater gains in facilitating rich classroom dialogue (and their students achieved higher learning gains) compared to teachers who engaged in traditional video reflection without AI support. These findings echo broader research suggesting that well-designed learning analytics tools can function as a "conversation partner" for teachers' reflection, helping them notice patterns they might otherwise miss and prompting data-informed discussions about teaching (Chen, Clarke, & Resnick, 2015). In sum, by situating AI within the DIKW framework, the presenters highlighted how AI can play a role at each level of the hierarchy – not only handling data and information processing, but also enabling knowledge-building and even the cultivation of professional wisdom through dialogue and reflection.

## 4. AI Tools and Applications: Design Principles and Implementations

**Triadic Human–AI–Expert Networks**
Moving from theory to design, participants presented several novel AI tools and models aimed at enriching educational dialogue. Li Kexin (University of Hong Kong) introduced a comprehensive model for integrating AI into teacher professional development (PD), which she called the "Triadic Synergy Model." This model deliberately interweaves three essential triads in the PD ecosystem:

1. **Participants Triad:** Teachers (as the agents of growth), human experts or mentors (providing experience-based guidance), and an AI agent (offering scalable support and resources).
2. **AI Affordances Triad:** Different functional roles for AI, including a visualization-based learning analytics system (to make dialogue patterns visible, as discussed above), a personalized chatbot (to provide on-demand coaching or feedback), and an explainable AI component for transparent assessment (ensuring teachers can understand AI-generated suggestions or evaluations).

3. **PD Design Triad:** Grounding the PD in research-informed practices (e.g. using design-based research principles), involving experts in co-designing activities, and being practice-driven through iterative pilot studies in real classrooms.

By aligning these three triads, Li's model creates dynamic teacher–expert–AI partnerships rather than a simple human–tool interaction. The rationale is that each element compensates and reinforces the others: AI can scale up support and provide data-driven insights; human experts ensure that the guidance is pedagogically sound and context-sensitive; and teachers remain central as reflective practitioners implementing and testing new strategies. An example scenario from Li's work involved a teacher who autonomously selects and uploads a classroom video to an AI coaching tool. The AI system then analyzes the teacher's class (via transcripts or live data) and offers feedback on their use of dialogic talk moves (like asking open-ended questions or prompting student elaboration). Afterward, the human mentor and teacher would discuss these AI observations, interpret them, and decide on next steps for improvement. This synergy addresses a common challenge in PD – teachers often struggle to recognize and differentiate effective talk moves in practice, sometimes adopting dialogic strategies only performatively (superficially) rather than purposefully. The combined support of expert mentors and AI feedback in Li's model helped teachers more genuinely internalize dialogic techniques and use them adaptively, rather than as a checklist. Overall, the Triadic Synergy Model exemplifies a "human-in-the-loop" approach to AI integration: AI is leveraged for what it does best (data crunching at scale, personalization, rapid feedback), but final interpretation and pedagogical decision-making remain with humans. This design philosophy was echoed throughout the workshop as a guiding principle for AI in education.

**Dialogue Training and Skill Development**
Another exciting frontier discussed was the use of AI to train and enhance students' and teachers' dialogue skills. Christa Asterhan and Avivit Arvatz (Hebrew University of Jerusalem) presented a project called PRODIGY, which proposes an integrative framework for defining and cultivating "productive dialogue." Drawing from cross-disciplinary research in science communication, humanistic psychology, and education, they identified four key dimensions of productive dialogue: (1) Content, i.e. the quality of knowledge co-construction and reasoning; (2) Interpersonal rapport, referring to the relational dynamics like respect, sharing and empathy; (3) Perspective-taking & listening, indicating how well participants attempt to understand and respond to each other's ideas; and (4) Intellectual humility and Integrity, meaning ethical participation and being trust-worthy in dialogue. Together, these dimensions provide a rich definition of productive dialogue for learning purposes, as well as an objective in itself. The PRODIGY project leverages generative AI as conversational agents for dialogue skill training. For instance, the team has developed AI-driven practice scenarios where students (or teachers in training) can engage in a conversation with an AI conversational agent that simulates a discussion partner that always takes the opposite stance on an issue. The AI-driven agent is programmed to exhibit certain dialogue behaviors and acts according to one of two different rhetorical styles: disputative and dismissive to train participants in dealing with real-life low-quality discussions, or deliberative and in accordance with Prodigy to experience and model "productive dialogue". Learners receive immediate feedback on their own dialogic performance from the system, which is provided according to the four PRODIGY dimensions, such as, for example, pointing out missed opportunities to ask a clarifying question (Content dimension) or instances where their response did not acknowledge the partner's perspective (Perspective-taking dimension). The underlying idea is that productive dialogue performance with a disagreeing partner on socio-political, "hot" topics can be improved through deliberate practice, much like writing or mathematics. An AI-driven conversational agent provides a safe, controlled environment for that practice. Generative AI is particularly suited for this because it can flexibly take on different roles or viewpoints in dialogue on current socio-political topics, allowing trainees to experience a range of conversational situations with fellow non-experts. Asterhan and Arvatz reported initial results indicating that learners who used the AI-mediated dialogue practice showed improvements in specific performance indicators (such as formulating probing questions and more actively building on others' ideas). This suggests that AI could serve as a valuable "dialogue simulator" or conversational coach in education – a novel application of AI beyond its typical content tutoring role.

Taking a different approach to dialogue training, Julie Cohen (University of Virginia) described the use of mixed-reality simulations that blend AI with human elements for teacher training in dialogic teaching practices. In her system, pre-service teachers enter a virtual classroom environment where some "students" are AI-driven avatars and others are controlled by live actors. The teachers practice leading discussions or facilitating group work in this simulated classroom, and the AI components can be tuned to exhibit certain behaviors (like a student avatar who always dominates the discussion, or one who never speaks unless prompted). This allows teacher candidates to practice strategies for inclusive and equitable dialogue – for example, how to draw a quiet student into the conversation or how to manage an overly dominant student – and receive immediate feedback. Cohen highlighted a fundamental tension that her work grapples with: Can relational, human-centered teaching skills be developed through interactions with AI-driven avatars? Some critics argue that pedagogical dialogue is inherently a human relational practice that might not transfer from a simulation to a real classroom. Cohen's research is exploring this question by comparing teachers' performance in real classrooms before and after training with the mixed-reality simulation. Early findings suggest that while simulations cannot replicate all nuances of human interaction, they do help teachers gain confidence and a repertoire of tactics (like specific question prompts or body language cues) for dialogic instruction. The key, Cohen noted, is to ensure trainees reflect on the experience and connect it to human relationships – for instance, discussing how an AI avatar's responses might differ from a real student's feelings. This line of work illustrates how AI, combined with controlled human input, can create "practice spaces" for educators to hone the art of dialogue in teaching.

**Automated Assessment and Coaching**

The workshop also featured AI applications aimed at assessing dialogue and providing coaching to educators in real classrooms. Lindsay Clare Matsumura (University of Pittsburgh) presented an innovative use of ChatGPT (a large language model) to assist teacher coaches. In her project, transcripts of classroom discussions are fed to ChatGPT with instructions to identify "coachable moments" – segments of the discussion where a teacher did something that could be improved or where a significant student idea emerged that the teacher didn't fully capitalize on. Essentially, the AI scans lengthy transcripts and flags portions that might be worthy of a coaching conversation. Matsumura reported that in preliminary evaluations, over half of the segments flagged by the AI matched those that expert human coaches would also have selected, and about a quarter of the AI-selected moments were near-misses (meaning they were pedagogically relevant, but from less optimal parts of the lesson). This level of overlap is promising, suggesting that AI could help human coaches save time by quickly surfacing potential discussion points. However, Matsumura also noted that the AI sometimes misses the deeper context – for instance, it might flag a moment as suboptimal based purely on transcript text, whereas a human coach knows from classroom context that the teacher's move was appropriate. Therefore, in practice the AI's suggestions would not replace a coach's judgment but serve as a starting point. One could imagine an AI assistant that, after a class, provides a teacher (or their coach) with a shortlist of "interesting moments" to review, along with reasoning for why they might be significant (e.g. "Here, three students built on each other's idea but the teacher changed the topic – this could be a missed opportunity to probe deeper"). Such tools embody the augmentative role of AI – enhancing human capacity by sifting through data and providing preliminary analyses, thereby freeing educators to focus on interpretation and action.

In a related vein, TJ McKenna (Boston University) demonstrated the NGSS AI Curriculum Coach, an AI system designed to help teachers implement inquiry-based science lessons aligned with Next Generation Science Standards (NGSS). The system uses a Retrieval-Augmented Generation (RAG) architecture: it combines a generative AI (like GPT) with a domain-specific knowledge base (in this case, a graph of science curriculum content and pedagogical strategies). The AI "coach" interacts with teachers through a chat interface, essentially holding a dialogue about the teacher's lesson planning or classroom moves. Importantly, McKenna emphasized that the AI coach is built to use Socratic questioning and reflective prompts, rather than simply telling the teacher what to do. For example, if a teacher asks the AI how to get students to engage in argumentation from evidence, the AI might respond with a question: "What opportunities for student talk do you see in this lab activity?", followed by suggestions drawn from the NGSS knowledge base (like, "Consider asking students to debate which experimental result best supports the hypothesis"). This approach models the kind of responsive, non-prescriptive dialogue that teachers themselves are encouraged to use with students. McKenna reported positive feedback from teachers piloting the AI coach – many found that it pushed them to reflect more deeply on their practice, as if "thinking with a partner," rather than just receiving canned advice. One design lesson from this project is the value of retrieving trusted curricular knowledge: by grounding the AI's responses in a vetted knowledge graph, the system ensured that its suggestions were aligned with sound pedagogy and accurate science, mitigating the risk of AI "hallucinations." It also enhanced transparency, since teachers could ask the AI to show the source or rationale behind a suggestion (e.g. citing a specific NGSS principle), thereby increasing trust in the tool. This aligns with a broader trend in AI-in-education towards explainable AI, where systems are designed to make their reasoning visible to users.

**Coding and Analysis Tools**

Several participants demonstrated AI-powered tools for coding, analyzing, and visualizing dialogue – essentially research and teacher-support tools that handle the heavy lifting of discourse analysis. As mentioned earlier, Gaowei Chen and Pengjin Wang showcased the Classroom Discourse Analyzer (CDA) which provides rich visualizations of classroom talk. Beyond the CDA, another tool highlighted was the Tech-SEDA analyzer, used by multiple teams. Tech-SEDA stands for Technology-enhanced

Systematic Educational Dialogue Analysis, a framework that extends a prior coding scheme (SEDA) with additional categories relevant in tech-mediated environments. During the workshop, Dongkeun Han (University of Cambridge) described how he combined Tech-SEDA coding with multimodal analysis techniques to study videoconference lessons. His tool could, for example, synchronize dialogue transcripts with video and gesture data, allowing analysis of not just what was said, but how participants' facial expressions or hand gestures contributed to the dialogue in an online setting. The AI assisted by automatically tagging certain features (like long pauses, overlapping speech, or use of specific keywords related to reasoning), which helped in identifying segments for deeper human analysis. This example shows how AI is accelerating the researcher's ability to analyze educational dialogues at scale – a task that traditionally required painstaking manual coding. By making dialogue features more readily quantifiable, such tools enable researchers and teacher-educators to quickly spot patterns (e.g., "Students in the experimental group showed more instances of exploratory talk after the intervention") and to provide teachers with feedback grounded in evidence.

An important aspect raised was explainability and teacher trust in these analytical tools. Chen presented research indicating that when teachers are given explainable AI outputs – for example, an AI coding of their classroom discussion accompanied by plain-language justifications or examples – their trust and acceptance of the technology increases. This addresses the "black box" concern: teachers are rightly wary of AI analyses that mysteriously label their teaching as, say, 40% dialogic without any clarity on how or why. One solution demonstrated was an interface where a teacher can click on an AI-generated code (like "critical question") and see the exact moment in the transcript that led to that code, with an explanation (e.g., "We identified this as a critical-thinking question because the teacher asked students to provide reasoning for their answers"). By involving teachers in the interpretation of the data, the tool becomes a springboard for reflective dialogue about practice, rather than an opaque judgment. This resonates with broader best practices in learning analytics and AI design: transparency and user control greatly enhance the utility of these systems in educational contexts.

## 5. Empirical Findings and Cross-Cultural Insights

**Technology-Mediated Dialogue Environments**
The workshop included rich empirical studies examining how AI and digital tools are mediating dialogue in diverse educational settings. Dongkeun Han (University of Cambridge) presented findings from a design-based research project with teachers in the UK and Mexico, focusing on the use of videoconferencing platforms for dialogic teaching. With the rise of remote and hybrid learning, videoconference tools (like Zoom or Teams) have become common classrooms, but their impact on dialogue is not fully understood. Han's study found that these platforms can indeed facilitate complex, multi-layered dialogues – for instance, allowing parallel interactions in the chat while a verbal discussion happens, or enabling students to bring digital resources into the conversation (screen sharing, virtual whiteboards, etc.). Such affordances can enrich dialogue by adding multimodal elements: text, visuals, and gestures augment spoken words. Interestingly, teachers in the study reported that quiet students sometimes found their voice in the chat or through reaction emojis, contributing ideas they might not voice aloud. However, Han also observed that effective dialogic teaching over video requires more explicit modeling of norms and strategies. Unlike a physical classroom where norms of turn-taking or active listening might be more tacit, in a Zoom class teachers had to deliberately demonstrate how to engage (e.g., explicitly encouraging students to use the "raise hand" feature, or modeling responses like "Let me build on what Maria said…"). The implication is that technology-mediated dialogue is feasible and can even be enriched by new modes, but teachers must adapt and guide students to maintain quality dialogue in these environments.

**Student Engagement and Critical Thinking**

How does AI influence student engagement and thinking during dialogue? Several studies tackled this question. Henrik Tjønn (University of Oslo) examined Norwegian middle school classrooms where students interacted with AI chatbots as part of their learning activities. A key finding in the first phase of the study was that when pairs of students engaged with a single AI chatbot together, it spurred deeper discussion and reflection than when students interacted one-on-one with the AI. In a paired scenario, students often discussed the AI's responses with each other – sometimes critiquing it, sometimes building on it – which effectively turned the AI into a trialogue partner (two students + AI) instead of a dyadic student-AI conversation. This led to more critical thinking for many of the students, as they were prompted to explain their reasoning to their peer especially when the AI gave an unexpected or confusing answer. By contrast, individual students using the chatbot tended to accept the AI's feedback more passively or only engage in a shallow Q&A. Tjønn's and colleagues work underscores the importance of social contexts in AI-mediated learning: even when AI is involved, arranging collaborative use (students working in pairs or groups with the AI as a joint resource) can preserve the benefits of peer dialogue and reduce the risk of over-reliance on the AI's answers. However, he also found that teachers needed to model how to have a meaningful dialogue with AI. Students who were simply given a chatbot often asked trivial questions or got stuck in unproductive exchanges, whereas a teacher modeling a few prompts (e.g., "Ask the AI why it gave that answer" or "Let's critique the AI's suggestion together") made a big difference in the quality of subsequent student–AI interactions. This suggests that dialogic pedagogy now extends to guiding how students converse with AI tools, treating it as a new genre of classroom talk. The students need first to learn about AI as tools and infrastructure, then use it to work with problems that AI can augment the learning processes, and lastly learn domain specific knowledge through AI tools that is tailored for such goals and aims. It's the three types of processes (about, with and through) that can create productive learning trajectories for students in and across knowledge domains.

**Emma Hayward (University of Cambridge) offered insights from a rather different context: Vocational Education.** She described various collaborative projects in which vocational students (in performing arts, gaming design and film) used cutting-edge AI technologies like motion capture and generative media tools. For example, students used both inertial sensors and AI-based motion capture to create digital dance performances and 3D gaming avatars, which required them to discuss both the artistic content and the functioning of the tools. Hayward observed that these projects led to remarkable student engagement through multimodal dialogue. Students were actively talking about their project goals, building on each others' ideas, troubleshooting how to get the AI tool to do what they wanted, and reflecting on the outcomes – often all at once, using a blend of technical language and creative expression. The technology served as a kind of boundary object that sparked conversation: students had to negotiate meanings (e.g., what does a "good" digital dance look like?), explain technical or aesthetic issues to people from other disciplines, and jointly make decisions in order to achieve their visions. Hayward noted one caution: the novelty of AI can initially distract from learning goals, but with teacher guidance to keep the dialogue focused ("How does this tool help us explore the concept we're studying?"), the AI became an amplifier of inquiry. This case shows that AI in vocational settings can open up new forms of multimodal, collaborative, hands-on dialogue – where talking, creating, and using technology can be deeply intertwined.

**Argumentation and Literary Analysis**

Two particularly thought-provoking contributions examined AI's role in higher-order thinking tasks like argumentation and literary interpretation. Michele Flammia (Independent Scholar, Italy) presented the Chat-With-CAT project, a cross-disciplinary, multi-institutional research initiative (Montclair State University, Iowa State University) aimed at creating an AI-supported system to help students develop their argumentative writing skills.

The project integrates analytical frameworks such as the Argumentation Rating Tool (ART) and the Rational Force Model (RFM), starting from the identification of chains of

reasoning—that is, the reconstruction of the multiple reasoning steps present in students' essays. Students will write an argumentative essay and then interact with an AI Computerized Automated Tutor "CAT," a chatbot with a defined personality. The CAT will provide feedback, asking students to clarify their arguments, provide and evaluate reasons and evidence, and consider counterarguments.

The project is still in its early stages, but it has the potential to help students reflect more deeply on their reasoning. However, it also raises important epistemological questions: when we ask an AI system to evaluate a student's argumentative response to an open-ended, controversial issue, a fundamental question arises: Do we assume that a single objective answer exists? Or do we believe the answer is co-constructed by the group through dialogue and the interplay of diverse perspectives?

Flammia argues that these questions require a clear answer before any AI support system can be meaningfully developed. Although it may seem like a purely philosophical concern, it has direct pedagogical implications: it influences the teacher's role in dialogic facilitation, shaping the degree of intervention, the management of disagreement, and the balance between epistemic authority and openness.

On the topic of literary analysis, Helen Lehndorf shared observations from literature classrooms where students used AI (like GPT-based tools) to generate ideas or even short analytical paragraphs about novels and poems. She characterized the AI as a potential "third voice" in the classroom dialogue. On one hand, AI could enhance ideation: students who felt stuck or overwhelmed by a difficult text used the AI to get a few interpretive angles, which then jump-started rich classroom discussions. In one case, an AI-generated interpretation of a poem was presented to the class (with students fully aware of its source), and students debated its merit – effectively treating the AI's take as just another perspective to consider. This use of AI as a brainstorming partner helped lower-achieving students contribute more, as they had something to build on or critique. On the other hand, Lehndorf warned of subtle negative effects: students who relied heavily on AI formulations sometimes mirrored the AI's language in their essays, leading to a certain uniformity and a missed opportunity to develop their own voice. Moreover, she noticed a performance gap widening – higher-achieving students used AI outputs as a springboard (often critically), while some lower-achieving students tended to copy or lightly rephrase the AI's words, reinforcing superficial learning. These findings echo concerns in recent research that overusing generative AI can impede the development of original writing skills and critical analysis. The implication for pedagogy is that if AI tools are to be used in humanities classrooms, teachers should actively design activities where AI outputs are treated as raw material for human interpretation, not final answers. For example, comparing an AI-generated analysis with a student-generated one and discussing differences can make the exercise about evaluation and critique, thus keeping the dialogue authentic and focused on student thinking.

## 6. Critical Issues and Pedagogical Tensions

While the applications above show great promise, the workshop participants also grappled with critical issues and tensions that arise from bringing AI into educational dialogue. These discussions often cut across the specific projects, reflecting shared concerns about the broader impact on pedagogy and learning.

**The Agency Question:** Who Is Learning?

A recurring question was "Who is doing the learning or thinking when AI is involved?" This speaks to the issue of student agency and cognitive engagement. During the workshop, participants shared anecdotal evidence: for instance, a teacher noticed that students using an AI-based tutor became adept at quickly getting answers but showed less improvement in underlying conceptual understanding compared to students who learned through peer discussion. The group agreed that maintaining student agency is paramount – AI should not be doing the intellectual heavy lifting all the time. One strategy suggested was to design AI activities that require students to reflect or explain in their own words. For example, if an AI provides a step in a math solution, the student might be prompted to explain why that step is valid before seeing the next step. This way, the student remains an active sense-maker. The tension here is clear: AI can easily present information or solve problems (often faster than a human can), but if we always let it do so, the student's role in knowledge construction diminishes. Thus, the workshop emphasized the importance of keeping students in the driver's seat of dialogue, using AI as a GPS rather than an autopilot.

**Ingvill Rasmussen (University of Oslo) highlighted a related issue at the classroom level:** the challenge of maintaining collective dialogue in the era of personalized AI. Many AI tools – particularly adaptive learning systems or chatbots – deliver highly individualized interactions. Rasmussen noted that even in cases where these tools improved individual student outcomes (like better explanations or faster problem solving), teachers observed a decline in whole-class discussions and collaborative sense-making. If each student is working with their own AI assistant, the traditional class dialogue (where ideas are shared and built upon in a group) can suffer. This raises a critical design question: How can we integrate AI in a way that supports community knowledge-building, not just isolated learner experiences? Some ideas included using AI to summarize or bring together contributions from different students to spark a group dialogue, or having periods in a lesson where everyone "unplugs" and engages in face-to-face discussion informed by their prior AI-assisted work. The consensus was that educators and technologists must be mindful of preserving spaces for dialogic interdependence – the process where learners learn from each other, not only from an AI, thereby developing skills like argumentation, listening to diverse viewpoints, and co-constructing knowledge.

**Equity and Access Concerns**

Participants also raised strong concerns about equity and inclusion in the context of AI-supported dialogue. One issue is the uneven engagement with AI among students.

Helen Lehndorf observed that in her classrooms, high-achieving students tend to use AI tools more actively and effectively, treating them as extensions of their curiosity, whereas some lower-achieving or more anxious students feel overwhelmed or sidelined by the AI. For example, a confident student might use a chatbot to explore an advanced idea and then bring that insight to class discussion, boosting their already strong performance, while a struggling student might either avoid using the AI (due to fear or confusion) or use it uncritically (thus not really learning, just echoing the AI). This dynamic could exacerbate existing achievement gaps if not addressed. The group suggested that teachers need to pay special attention to how different students engage with AI, potentially offering scaffolding or alternatives for those who are less comfortable. Also, tools should be designed with adaptive support – perhaps an AI that can detect if a student is stuck or just copying and then switch mode to ask more guiding questions.

Avivit Arvatz passionately argued for "human-in-the-loop" solutions that empower rather than marginalize learners. She cautioned that if AI is used naively, it might reinforce a deficit perspective (e.g., allowing for disrespectful and impolite talk to the AI that can diffuse into human-to-human interaction ). Instead, an inclusive approach would involve students in understanding and even developing the AI tools – for instance, having students critique the AI's performance or suggest improvements, which can demystify the technology and give them ownership. Moreover, equitable AI in education also means addressing language, cultural, and socioeconomic biases. If a dialogue system is only trained on data from Western classrooms, its interaction style might not resonate in other cultural contexts, or it may lack sensitivity to multilingual settings. Participants noted ongoing efforts, such as translating dialogue prompts and incorporating culturally diverse scenarios in AI training, but clearly, more work (and research) is needed to ensure AI does not widen educational inequities. An encouraging example came from one project where students in under-resourced schools were taught to use an AI translator chatbot to communicate with peers in another country – this empowered them to participate in a cross-cultural dialogue that would otherwise be inaccessible due to the language barrier.

Such cases show AI's potential for inclusion if used thoughtfully. The overarching recommendation was that equity must be a design principle from the start when creating AI for educational dialogue, and continuous monitoring for disparate impacts should be integral to any deployment.

**Authenticity and Skill Development**

Another tension discussed was the authenticity of AI-mediated dialogue and its impact on developing genuine communication skills. Heather Hill (Harvard Graduate School of Education) voiced skepticism about student-to-AI dialogue as a substitute for student-to-student or student-to-teacher dialogue. She reminded the group that dialogic instruction, as understood in educational research, is fundamentally a relational, human practice – it's not just exchange of information, but also about empathy, trust, and personal connection. An AI might be very good at the cognitive aspects (asking thought-provoking questions, giving feedback), but it lacks the human emotions and responsiveness that build real social support. Hill questioned whether practicing with AI could truly inculcate the same dispositions and skills that practicing with humans does. For example, will a student learn to listen attentively and respond empathetically if the partner is an AI that has no feelings? Some feared that extended interaction with unemotive AI partners might even habituate students to a conversation style that is more transactional and less attuned to subtle interpersonal cues – potentially reducing skills like empathy or deep listening when the student later engages with humans. This concern echoes others raised in media about AI chatbot usage affecting human interaction habits (Turkle, 2015). Dongkeun Han added that students might develop conversational habits with AI (like always expecting a quick answer, or not pausing to think since the AI fills silences) that carry over negatively into human dialogues. The group acknowledged these as open questions needing further study. As a precaution, some suggested that AI-facilitated activities should always be complemented by reflection sessions focusing on the human element: e.g., after an AI role-play debate, students could discuss how it felt compared to debating a classmate, and identify what emotional or collaborative aspects were missing. By being explicit, educators can help students be more meta-cognitive about what good dialogue entails, beyond what the AI can provide.

**Conceptual issues for discussion**

Finally, the workshop touched on what might be called conceptual issues – concerns that over-use of AI could displace or erode fundamental human dialogic processes and values. Yifat Kolikant (Hebrew University of Jerusalem) made an interesting argument here: she suggested that AI might best be used in contexts where rich dialogue is unlikely to occur naturally, rather than in those where it traditionally thrives. For example, in classrooms or topics known to generate little discussion (perhaps due to controversy, discomfort, or lack of knowledge), an AI agent could be introduced to stimulate or mediate dialogue. One could imagine an AI that moderates a discussion on a polarizing social issue, ensuring all voices are heard and providing fact-checks – essentially seeding a dialogic process where none might have existed. However, Kolikant warned strongly against allowing AI into the core spaces of human-to-human exploratory talk. If students can discuss openly and critically among themselves or with a teacher, that is gold-standard dialogic learning and should not be handed off to AI mediation. The risk is that if teachers become reliant on AI to run discussions or if students default to asking AI instead of a peer, we could see a diminishment of the culture of dialogic inquiry that educators have worked hard to establish. The consensus was that AI's role should be strategic and limited – used as a scaffold or enhancer in scenarios of need, but never to replace the essential human dialogue that builds classroom community, trust, and critical citizenship. In other words, keep the heart of dialogue human. Participants invoked an analogy: just as calculators are used in math class but we still teach mental arithmetic to develop number sense, AI can be used in dialogue-rich classes but we must still cultivate students' ability and willingness to dialogue with other humans, which is foundational to democratic education.

## 7. Methodological Innovations and Coding Frameworks

In addition to pedagogical insights, the workshop showcased innovative methodologies for studying and assessing dialogues in the age of AI.

**New Analytical Approaches**
Sten Ludvigsen argue that researchers could exploring new frameworks like mechanical philosophy to conduct multilevel analyses of learning environments. This involves methods that can connect interaction data (from chat transcripts, video, AI logs, etc.) to larger systemic factors (like school practices or technology infrastructure). One example discussed was using network analysis to map all interactions in a tech-rich classroom: human–human, human–AI, and even AI–AI (e.g., one AI system feeding data to another). By visualizing such networks, researchers can identify central nodes (perhaps a particular student who mediates between others and the AI) or bottlenecks (points where information flow is restricted). This approach, albeit data-intensive, can yield insights into how knowledge and authority circulate in an AI-enhanced class. Another approach mentioned was temporal analysis using time-series methods: looking at how the introduction of an AI tool changes discourse patterns over time. For instance, does classroom talk become more student-driven or more teacher-driven after a certain AI is adopted? Such analyses require sophisticated coding schemes and often AI assistance to handle the coding, but they push the field toward understanding causal and developmental dimensions of AI's impact.

Asterhan and Arvatz's work on operationalizing productive dialogue criteria (the PRODIGY framework) is also a methodological contribution. By translating often qualitative dialogue features into measurable indicators, they enable the use of AI both to simulate human dialogue (for training, as discussed) and to automatically assess dialogue quality. During the workshop, they shared how they defined metrics for things like listening: for example, if in a dialogue the second speaker's contributions frequently reference or paraphrase the first speaker's points, that's evidence of good listening and perspective-taking. Such patterns can be detected via text analysis or even audio analysis (identifying overlaps or backchannel signals). Developing these metrics is challenging – it requires balancing nuance with consistency – but it's a step toward the use of AI in dialogue-based interventions. If successful, teachers could one day get machine feedback providing indicators of progress in dialogic teaching. However, participants also cautioned that not everything important is easily measurable; the risk is creating a new reductive metric (e.g., "dialogic score") that oversimplifies. Thus, these analytical innovations are best seen as tools to assist human judgment, not replace it, a point echoed throughout the event.

**Tech-SEDA Framework Applications**

The Technology-enhanced SEDA (Tech-SEDA) framework received significant attention as a versatile tool for comparative research. Originally, SEDA (Hennessy et al., 2016) was a coding scheme to characterize classroom talk moves and dialogue types in a systematic way. The "Tech-" extension updates this for the digital age, adding codes for things like tool use, multi-modal references, or new forms of teacher orchestration in tech-mediated settings. At the workshop, multiple teams reported using Tech-SEDA in different countries and contexts, enabling some comparative insights. For example, Han's study in the UK vs. Mexico found that certain dialogic moves (like students asking questions to each other) were more frequent in the UK classrooms, whereas Mexican classrooms saw more teacher prompts for elaboration – a difference potentially cultural or tied to teacher training. By using a common framework, researchers could discuss whether such differences were amplified or reduced when AI tools were introduced. The general sentiment was that having shared frameworks like Tech-SEDA can foster international collaboration, as data can be shared and compared more easily. Participants expressed interest in building a shared corpus of coded dialogues from various AI-in-education projects. This could accelerate learning about what dialogic patterns are universally beneficial versus those that are context-dependent. It could also feed into AI development: a large, labeled dataset of educational dialogues could be used to train AI that better understands pedagogical contexts (for instance, a chatbot that knows when to ask an open question versus when to give direct feedback). The workshop's discussions hinted at an emerging network of dialogue researchers keen on pooling their methodological tools and data – a collaborative trend that could significantly advance the field.

**Explainable AI in Education**

The theme of Explainable AI (XAI) came up not only in tool design but as a research direction in itself. Gaowei Chen's presentation provided evidence that making AI decisions transparent improves teacher uptake. Building on that, participants debated how to systematically study and improve explainability in educational AI. Some proposed developing standard evaluation protocols for educational XAI: for instance, measuring how well teachers or students understand an AI system's outputs, and whether that understanding leads to better decisions or learning outcomes. One example was an experiment where teachers were divided into two groups when using an AI grading assistant – one group got only scores from the AI, another group got scores plus explanations for each score. The hypothesis was that the latter group would have more trust and be more willing to act on the AI's feedback. Preliminary reports indicated this was true, but interestingly, those teachers also spent more time discussing and sometimes questioning the AI's rationale, which became a professional learning opportunity in itself. This suggests that XAI not only increases trust but can provoke fruitful dialogue about practice (e.g., "The AI thought my question was low-level; do I agree? Maybe I should refine it."). Researchers see potential in designing AI systems that deliberately encourage this reflective dialogue. For instance, one idea was an AI that occasionally admits uncertainty or offers two possible interpretations of a student's answer, and asks the teacher to weigh in – essentially prompting teachers to do a bit of formative assessment alongside the AI. Studying such interactions could reveal how humans and AI can collaboratively reason about educational data. In sum, making AI more explainable is not just a matter of ethics or user-friendliness; it's increasingly viewed as a feature that can enhance learning if done in a dialogic way, inviting users to engage with the AI's reasoning process.

## 8. Design Principles for AI in Educational Dialogue

Throughout the presentations and discussions, several design principles emerged as guiding values for developing and implementing AI in educational dialogue contexts. These principles are informed by both successes and pitfalls identified in research and practice.

**Human-in-the-Loop Approaches**

The foremost principle is to keep humans in the loop at critical junctures of AI systems. Whether it's in automated teaching assistants, dialogue simulators, or analytics tools, participants agreed that AI should not function autonomously in high-stakes decision-making or interaction without human oversight. Li Kexin's Triadic Synergy Model is a prime example of operationalizing this principle: the AI is deliberately positioned as a mediator and supporter, never the sole source of guidance. In practical terms, this might mean AI provides a recommendation but a teacher or student must confirm or modify it; or an AI facilitates a discussion but a teacher sets it up and debriefs it. Designing systems this way helps ensure that the use of AI amplifies human agency instead of diminishing it. It also aligns with emerging ethical frameworks that call for human control and accountability in AI deployment in classrooms. At the workshop, some shared design tactics included: requiring a "human check" for any AI-generated content that will be shared with a class, allowing users to easily correct the AI or provide feedback to it (making the AI learn from the human, in a sense), and dashboards where a teacher can see an overview of all AI-student interactions to monitor for issues. The human-in-the-loop mindset is essentially a safeguard against abdication of educational responsibility to algorithms – it keeps teachers as orchestrators of learning and students as active learners, with AI as an empowering tool under their guidance.

**Domain-Specific vs. Generative AI**

A nuanced design consideration discussed was the choice between domain-specific AI systems and more general generative AI (like large language models) for educational purposes. Sten Ludvigsen drew an important distinction: prior to around 2022, many educational AI systems were expert/knowledge-based – built on curated domain knowledge and predefined rules – whereas newer systems like GPT-4 are stochastic generative models that predict text without guaranteed factual accuracy. Each approach has implications. Domain-specific systems (for example, a physics tutor that has a hard-coded database of physics principles) can offer reliable, accurate guidance and align closely with curriculum standards, but they may lack flexibility or the ability to handle unexpected questions. Generative models are extremely flexible and can converse on almost any topic, often appearing remarkably "intelligent," but they can produce incorrect or nonsensical information and require external validation. The consensus was that educational design should favor a hybrid approach: use generative AI's natural language capabilities and adaptability, but constrain or augment it with domain-specific knowledge to ensure quality. McKenna's RAG-based coach exemplifies this by tying a generative AI to a vetted curriculum graph. Another example discussed was having an AI writing assistant for students that uses a large language model to give feedback on style and coherence, but also checks content against a knowledge base or sources to verify facts (thus teaching students to be mindful of evidence). The group cautioned against deploying raw generative AI in high-stakes learning without such safeguards. As Ludvigsen bluntly put it during the meeting, "LLMs are not knowledge systems, they're probability models – treat them accordingly." Designers should implement features like fact-checkers, citations, or simple "Are you sure?" prompts when a student might be about to trust an AI's dubious answer. By understanding the strengths and limits of each AI paradigm, developers can create domain-aware generative tools that offer the best of both worlds.

**Scaffolding and Progressive Release**

Another principle emphasizes the idea of scaffolding AI usage and progressively releasing responsibility to learners. TJ McKenna and others advocated that AI tools, especially those intended for novices (be they students learning a subject or teachers learning a new pedagogy), should be designed to gradually shift control to the user as their competence grows. This mirrors the educational concept of the "gradual release of responsibility" model (Pearson & Gallagher, 1983) but now applied to human–AI collaboration. For instance, an AI might start a course by guiding a student very closely through solving problems (providing step-by-step hints). As the student improves, the AI could step back, giving only high-level hints or asking questions instead of giving answers. Finally, the AI might mostly observe, only intervening if it detects impasse or frustration. Such design ensures that learners don't become overly dependent on the AI and continue to develop their own skills and metacognition. In teacher-facing tools, a similar approach can be taken: early in a PD program, the AI coach might offer concrete examples and scripts for the teacher to try; later, it might just prompt the teacher to reflect without giving the answer, treating the teacher as a more equal partner. This progressive handover was illustrated by one participant who designed an AI for history essay writing: initially it gave students a detailed outline based on their thesis, later it just suggested a few key points, and by the end it would only remind the student to check their sources and argument structure, letting them do the rest. The ultimate goal is for the learner to internalize the guidance and not need the AI at all for that task – somewhat paradoxically, a successful educational AI might "work itself out of a job" for a given student. Workshop attendees agreed that this principle aligns AI use with sound pedagogy by preventing stagnation under constant support, instead promoting independence and mastery.

**Transparency and Accountability**

A final critical design principle is transparency, not just of AI algorithms (as discussed under XAI), but of roles, data usage, and goals – in short, accountability to the stakeholders (teachers, students, parents). Helen Lehndorf's Actor-Network Theory analysis underscored the importance of "traceable chains of reference" in AI-mediated learning: students and teachers should always be able to trace where information or feedback is coming from and how it was generated. This could be as simple as an AI tutor clearly indicating, "I used your last 3 answers to decide to give you this hint", or as broad as schools having a policy to inform students which of their interactions might be recorded or analyzed by an AI system. Transparency fosters trust and allows users to attribute success or failure appropriately (e.g., if an AI-made suggestion fails, understanding that it was the AI's idea, not their own misstep, which is an opportunity to learn and adjust). It also ties into ethical considerations like data privacy and informed consent – topics that, while not the main focus of this workshop, were acknowledged as ever more important as AI tools proliferate in classrooms. Effective AI integration must come with clear guidelines on what the AI will do, what data it will use, and how its recommendations should be evaluated. An example raised was a school that introduced an AI essay feedback tool: they held an orientation session for students explaining the tool's purpose, its limitations (it checks grammar and coherence but doesn't judge creativity), and emphasizing that final responsibility for the work lies with the student. Such practices ensure that accountability in learning remains with the humans – the AI is a means, not an actor with authority. Designing for transparency might reduce some efficiency or complexity (a fully "black box" AI might be more optimized, but inscrutable), yet participants strongly felt that in education, pedagogical and ethical imperatives outweigh the marginal gains of opaque efficiency. In summary, to be welcomed in the classroom, AI systems must behave in ways that are understandable and answerable to the educational community's values.

## 9. Future Research Directions and Collaborative Opportunities

The workshop concluded by identifying several areas where further research is needed, as well as opportunities for collaboration across institutions and disciplines. These future directions aim to address the gaps and challenges uncovered during the discussions.

**Methodological Needs**

**Longitudinal Impact Studies:** A clear need is for long-term studies on how sustained AI use affects students' dialogic skills and dispositions over time. For example, following a cohort of students who use AI support throughout high school to see how their critical thinking, collaboration, or communication skills evolve compared to those without AI. Do benefits persist? Are there delayed costs (e.g., a drop in creativity)? Longitudinal data would help parse out short-term novelty effects from lasting transformation, giving a more solid evidence base for policy. It would also inform the agency question: do students habituated to AI become more passive or, alternatively, more empowered as learners over years?

**Cross-Cultural Validation:** As multiple participants stressed, context matters. Tools and strategies that work in one educational culture may not directly translate to another. Therefore, cross-cultural studies of AI in dialogue – deploying the same intervention in schools across different countries or cultural settings – are needed to test generalizability. These can reveal cultural biases in AI systems (e.g., a discussion prompt generator might assume certain norms of debate that don't hold universally) and help adapt frameworks like Tech-SEDA to incorporate culturally specific dialogic practices. Such research promotes cultural responsiveness in AI design.

**Multimodal Analysis Methods:** The group highlighted the importance of developing methods to analyze multimodal dialogue (speech, text, gestures, drawings, etc.) in technology-rich environments. Traditional dialogue analysis often focuses on transcripts alone, but learning with AI can involve interacting with graphs, simulations, or physical devices, each contributing to the dialogue. New analytical techniques – possibly using machine learning for image or video analysis – are needed to capture the full spectrum of such interactions. This also ties into measuring engagement and collaboration more holistically (e.g., can we detect when a group is truly collaborating by looking at how they pass a tablet around or how often they all look at a shared screen?).

**Ethical Framework Development:** While not the central theme of each talk, ethical considerations underlie all of this work. Participants noted the urgency of creating guidelines for responsible AI integration in education. This includes data privacy standards, clarity on intellectual property of AI-generated student work, and strategies to prevent AI from amplifying biases or inequalities. Research is needed to develop and test such frameworks, potentially in partnership with organizations like UNESCO or governmental bodies (e.g., exploring how the UNESCO "AI and education: Guidance for policy-makers" document plays out in real school settings). Empirical ethics studies – such as surveying stakeholder attitudes or running controlled comparisons of different policy approaches – could inform best practices that balance innovation with student rights and well-being.

**Theoretical Developments**

The workshops show that cognitive, socio-cognitive and socio-cultural stances are strong within the learning theories when educational dialogue and AI are emphasised. They contribute with well-established methodologies and methods (such as coding schemas). Advanced statistical studies and different forms of qualitative analysis are used here. These methods and stances have already been adjusted and include units of analysis that capture AI as part of interactions and dialogues. This was demonstrated in many of the participants' presentations and can be seen in their publications. These three established stances also come with research design and design principles that can integrate AI as part of instructional sequences and learning trajectories.

In addition, suggestions for new conceptual ideas were put forward, such as **Rupert Wegerif's** provocative ideas. There is an opening for theoretical exploration of **Quantum Theory Metaphors** in dialogic education. While sounding abstract, this line of thought could push the boundaries of how we think about the non-linear, non-binary nature of dialogues that include AI. It encourages interdisciplinary scholarship bridging physics, philosophy, and education to ask new questions: Can concepts like superposition or entanglement metaphorically describe a state where a learner's thinking is "in two places at once" due to human–AI discourse? Could that yield useful insights or just poetic ones?

Research here would likely be conceptual and speculative at first, but could eventually inform how we design AI that participates in learning dialogues (e.g., an AI that can hold multiple perspective threads at once).

**Mechanical Philosophy Applications:** Sten Ludvigsen's call for robust multilevel analysis through a mechanistic lens suggests formulating new models of learning processes that explicitly include AI as components. For instance, one might attempt a formal model of a classroom as a complex system of interlocking "mechanisms" (some human, some AI, some material tools) and try to derive principles of how learning emerges from their interactions. This could integrate theories from the learning sciences with systems engineering. It's a tall order, but progress here could lead to theoretical frameworks that predict outcomes of certain designs – something that can currently only be discovered empirically.

**Actor-Network Theory Extensions:** Helen Lehndorf's approach invites further development of actor-network theory (ANT) in education to include AI actors. Traditional ANT analyses in education have looked at things like how textbooks, teachers, and students form networks that shape knowledge. Now we have AI agents in the mix. Future research could map these networks in detail and examine how power and agency are redistributed. For example, does introducing an AI tutor shift some authority away from the teacher node towards the software node, and with what consequences? Such studies blend qualitative observation with theoretical interpretation, and could yield rich narratives (e.g., case studies of a "network" of a classroom with AI and how an idea travels through it). They might also give designers insight into unintended effects (like the teacher feeling alienated when an AI gives students feedback directly).

**Technical Innovations**
**Improved Explainable AI:** On the technical front, participants saw a need for advancing explainable AI tailored for education. This includes developing algorithms that can not only make a decision or recommendation but also output pedagogically meaningful explanations. For example, an AI that assesses a student's essay could highlight specific sentences and link them to rubric criteria, or a math tutor AI could display the step-by-step logical path it used to solve a problem in a way a student can follow. Research here is inherently interdisciplinary: it requires AI experts working with educators to figure out what kinds of explanations truly aid learning (some might be too detailed, others too simplistic). Progress could transform AI from a mysterious oracle to a transparent mentor.

**Multimodal AI Integration:** A technical challenge mentioned was creating AI systems that can handle multiple modes of communication simultaneously. For instance, imagine a tutoring system that can watch a student's face for confusion, listen to their explanation, analyze their written work, and then intervene appropriately. We are still far from that level of integration. Research in this area overlaps with fields like computer vision (for gestures or facial expression recognition), speech analysis, and natural language processing, all synchronized in real time. Educational data (with privacy protections) could fuel this research – e.g., using classroom videos to train models that detect when a student's nonverbal cues contradict their spoken confidence. The payoff would be AI that better "understands" the full context of a learning interaction, thus responding more like a tuned-in human tutor who notices when a student furrows their brow.

**Domain-Specific AI Development:** Another future direction is building educationally-purposed AI models that go beyond generic LLMs. This might mean training language models on text corpora of educational dialogues, curriculum materials, and student-generated content, so that their style and knowledge is steeped in how teachers and students actually communicate. There is already movement in this direction (for example, customized models for programming education or for historical dialogue simulations), but more work is needed to cover different subjects, age groups, and languages. These models could be smaller and more efficient than massive general ones, making them easier to deploy in schools (perhaps even offline). The research question is whether such specialized models can achieve better educational outcomes than one-size-fits-all AI. Intuitively, an AI that "speaks education" – using age-appropriate vocabulary, explaining step by step, referencing common student misconceptions – would be more effective than a general chatbot. This needs empirical testing, and development to gather the right training data responsibly.

**Collaborative Research Networks**

Finally, the workshop itself exemplified the value of international collaboration, and participants identified several concrete areas for collaborative work:

**Shared coding frameworks and tools:** There was enthusiasm for forming a consortium to further develop and share tools like Tech-SEDA, CDA, and others, which would include pooling anonymized data to improve these tools' AI components. A common platform or repository could allow researchers globally to contribute data and get analyses back, accelerating findings for all.

**Cross-cultural studies:** As noted, running parallel studies in different countries could be facilitated by collaboration. For instance, a group of researchers could agree to use the same research design to test an AI tool in, say, five different countries, then share results. Collaborative grants (perhaps through OECD or EU funding for digital education) could support such efforts, ensuring diverse contexts are represented.

**Ethical guidelines development:** Bringing together experts in education, AI, law, and philosophy across institutions to hammer out ethical guidelines was seen as crucial. The group floated the idea of an "AI in Education Ethics Task Force" or a working group that could build on UNESCO's recommendations and produce a white paper or handbook for schools and companies. Input from multiple regions would ensure the guidelines have global relevance and legitimacy.

**Co-design of new tools:** Researchers and teachers from different countries could jointly design and pilot new AI tools, incorporating design-based research cycles. This not only shares workload and ideas, but ensures that tools are not tailored to an overly narrow context. For example, a collaborative project might design a dialogue agent that works in bilingual classrooms (with researchers from bilingual communities co-leading), benefiting all partners.

In essence, the workshop attendees recognized that the questions posed by AI in educational dialogue are bigger than any one team or discipline. By forming a network – effectively an international research community on educational dialogue and AI – they hope to leverage complementary expertise and contexts to advance knowledge and practice more rapidly. This spirit of open collaboration was one of the inspiring outcomes of the meeting, suggesting that as fast as AI is moving, the education research community is moving to keep pace together.

## 10. Conclusions and Implications

**The Workshop on Educational Dialogue:** Moving Thinking Forward (AI Strand) illuminated both the tremendous potential and the significant risks of integrating artificial intelligence into educational dialogue. AI tools, when thoughtfully designed and implemented, can indeed enhance classroom dialogue in multiple ways: they can visualize complex interactions to make discussion dynamics clearer; provide students and teachers with immediate feedback and personalized support; and offer new opportunities for practicing communication skills in simulated environments. The workshop's many examples – from AI coaches that engage teachers in reflective discourse, to chatbots that help students practice argumentation – demonstrate that, used wisely, AI can act as a catalyst for productive talk and a scaffold for developing higher-order thinking skills.

At the same time, the discussions underscored that these benefits do not come automatically or without caveats. There is a genuine danger that poorly implemented AI could undermine the fundamentally relational and human aspects of education. Authentic educational dialogue is not just about exchanging information; it is about building trust, listening, challenging and extending ideas, and often navigating ambiguity and disagreement. If AI tools encourage shortcutting of those processes – for instance, by giving quick answers that students accept uncritically, or by reducing face-to-face discussion in favor of individualized tutoring – then the very qualities that make dialogue so powerful for learning could be eroded. The group repeatedly emphasized vigilance against such outcomes and the need for ongoing research and reflection as AI use expands.

**Several key insights emerged as guiding principles moving forward>**

- **AI works best as a mediating tool within human networks and infrastructures, rather than as a replacement for human interaction.** The consensus was that AI should augment and amplify human dialogue, not take its place. For example, McKenna's curriculum coach and Li Kexin's triadic model both show AI embedded in a system that keeps teachers and experts central. In practice, this might mean using AI to spark a class debate or to assist a teacher's analysis, but still ensuring that students and teachers are actively engaging with each other around that AI contribution.
- **Equity and access must be priorities to prevent AI from exacerbating educational inequalities.** It was noted that tech innovations can sometimes widen gaps – those with more resources or savvy benefit most. Workshop participants stressed designing AI interventions with inclusive pedagogy in mind and providing support so all students (and schools) can leverage the tools. This includes considering multilingual and multicultural adaptability, as well as addressing the differential comfort levels students have with AI. AIs in education should be tested for unintended bias and differential impacts, and policies should aim to make transformative AI tools available broadly, not just to elite institutions.
- **Theoretical frameworks in education need updating to include AI's role, but without abandoning core dialogic principles.** As discussed, concepts like "dialogic space," agency, and scaffolding may require redefinition in light of AI, yet the essence of dialogic teaching – encouraging multiple perspectives, reasoned discourse, and collaborative meaning-making – remains crucial. Curriculum design and teacher training should evolve to integrate AI without losing sight of interpersonal values. In fact, several participants argued that as AI takes on more routine tasks, human educators should double down on cultivating the uniquely human aspects of dialogue (empathy, ethics, creativity). We may need new blends of frameworks – for instance, merging dialogic pedagogy with human–computer interaction models – to guide teachers in orchestrating classrooms where AI is present but human dialogue still thrives.

- **Design principles for AI in education should prioritize transparency, accountability, and human agency.** A recurring conclusion was that how we implement AI makes all the difference. If AI systems are transparent in operation, users can understand and critique them (building trust and avoiding over-reliance). If they are accountable – meaning they let humans make final calls and encourage user input – then the locus of control stays with the educators and learners. McKenna's approach of local-first deployment (where a school district maintains control of the AI tool and its data, rather than a remote company) was cited as a model that maintains human oversight and contextualization. Essentially, educational AI should conform to the pedagogy, not force pedagogy to adapt to a rigid AI.
- **Professional development and teacher support are critical:** educators need to experience and model dialogic practices with AI themselves. Teachers will be the linchpin of successful AI integration. As one insight highlighted, it's not enough to train teachers about dialogic strategies; PD programs should embody those strategies, with AI tools included in the process. McKenna's work showed that when teachers engaged in AI-mediated dialogic learning (for example, conversing with an AI about curriculum), they developed a more concrete understanding that they could transfer to their classroom. If teachers learn in a dialogic way – perhaps even co-designing AI use cases – they are more likely to cultivate that environment for students. Therefore, investment in teacher-centric design and training may yield the highest returns for moving the field forward.

An especially poignant reflection from the workshop was that the field of educational dialogue alongside AI is at a critical juncture. Decisions made in the next few years – by researchers, technology developers, educators, and policy-makers – will have profound implications for the future of learning, human interaction, and even democratic society. Educational dialogue is closely tied to how people learn to reason, to empathize, and to participate in civic discourse. As AI becomes more embedded in our communication, maintaining the health of dialogic education is both a pedagogical and a civic imperative. Will AI be used to foster more open, inclusive, and thoughtful conversations, or will it inadvertently stifle the human spirit of inquiry? The outcomes depend largely on the collaborative, wise efforts of the kind initiated at this workshop.

On a hopeful note, the international and interdisciplinary collaboration evident in Cambridge provides a strong foundation for tackling these challenges. By bringing together diverse expertise – from learning scientists and computer scientists to classroom practitioners – and by openly sharing findings, the community can ensure that AI integration is approached responsibly and creatively. The workshop is just one step in a continuing journey. The participants departed with new connections, a shared sense of purpose, and concrete plans for joint research and dialogue (both with each other and with their AI "partners"). Such collective engagement gives reason to be optimistic that the education sector can indeed harness AI's rapid advancements while preserving and even enhancing the humanistic, dialogic core of teaching and learning.

**Acknowledgments:** This workshop was co-organized by members of the Cambridge Educational Dialogue Research Group (CEDiR) at the University of Cambridge and Brigham Young University, with generous support from their institutions. The organising committee consisted of: Sara Hennessy (University of Cambridge), Rupert Higham (University College London), Bryant Jensen (Brigham Young University), Adam Lefstein (Hebrew University of Jerusalem) and Alison Twiner (University of Cambridge). We thank all participants for contributing their insights, research, and passionate perspectives. Their collaboration and critical dialogue over the three days have substantially advanced our understanding of AI's role in educational dialogue. We also acknowledge support staff and facilitators who ensured the workshop's smooth operation.

Any opinions, findings, or conclusions expressed here are those of the authors and do not necessarily reflect the views of the organizing institutions.